\documentclass[lettersize,journal]{IEEEtran}
\usepackage{amsmath,amsfonts}
\usepackage{algorithmic}
\usepackage{algorithm}
\usepackage{array}
\usepackage{textcomp}
\usepackage{stfloats}
\usepackage{url}
\usepackage{verbatim}
\usepackage{graphicx}
\usepackage{cite}
\usepackage{cite}
\usepackage{amsmath,amssymb,amsfonts}
\usepackage{algorithmic}
\usepackage{graphicx}
\usepackage{textcomp}
\usepackage{xcolor}
\usepackage{booktabs}
\usepackage{tabularx}
\usepackage{array}
\usepackage{hyperref}
\usepackage{caption}
\usepackage{adjustbox}
\usepackage{bm}
\usepackage{multirow}
\usepackage{tikz}
\usepackage{colortbl}
\usepackage{subcaption}
\begin{document}

\title{Frequency-Domain Compliance Assessment of Grid-Forming Devices}

\author{{Ambuj~Gupta},~\IEEEmembership{Student~Member,~IEEE}, {Muhammad~Sharjeel~Javaid},~\IEEEmembership{Member,~IEEE}, {Balarko~Chaudhuri},~\IEEEmembership{Fellow,~IEEE}, {Mark~O'Malley},~\IEEEmembership{Fellow,~IEEE}
\thanks{This work was funded by a Leverhulme International Professorship, grant reference [LIP-2020-002] and by the Engineering and Physical Sciences Research Council [EP/Y025946/1].}
\thanks{The authors are with the Control \& Power Group, Imperial College London, London, UK. (corresponding author e-mail: \href{mailto: a.gupta23@imperial.ac.uk}{a.gupta23@imperial.ac.uk}).}}



\maketitle

\begin{abstract}
Grid-ForMing Inverters (GFMIs) are expected to provide voltage stiffness to the grid. Explicitly, system operators (SOs) and regulators expect GFMIs to behave like a ``voltage source behind impedance (VSBI)'' in the (sub)-transient time frame. SOs assess this VSBI characteristic of GFMIs during compliance by defining a pass-fail time-domain criterion. This is done by evaluating the GFMIs' active (or reactive) power/current response to step changes in voltage phase (and magnitude) at its terminals. However, this approach is prone to errors due to poorly defined measurement specifications for very fast (less than a cycle) transients. To address this, this work proposes a compliance criterion for the VSBI characteristic of GFMIs in frequency-domain based on elements of the frequency-domain Jacobian. The compliance criterion is defined in terms of the minimum expected $P(s)/\theta(s)$ and $Q(s)/V(s)$ Bode plot characteristics across a specific frequency range. The equivalence between the time-domain and frequency-domain criteria is established. The proposed method is demonstrated by assessing the compliance of generic NLR (formerly NREL) GFMI models in PSCAD. Furthermore, the impact of GFMI compliance on the small-signal stability of the IEEE 39-bus bulk-power system is demonstrated.
\end{abstract}


\section{Introduction}
Regulators, system operators (SOs), and consortia worldwide have published technical specifications, functional requirements, and assessment criteria for grid-forming capabilities \cite{Gupta2026arxiv2602, eso2021gc0137, fingrid2023study, aemo2023voluntary, ERCOT2024AGS_ESR,miso2024gridforming, entsoe2024report}. Grid-ForMing inverters (GFMIs) are expected to behave as ``voltage source behind impedance (VSBI)'' in the sub-transient time frame during disturbances and in normal conditions, by maintaining a (nearly) constant internal voltage source (IVS) behind an impedance. To assess a GFMI's VSBI capability, most SOs expect a GFMI to respond naturally and instantaneously to changes in voltage angle (and magnitude) at its terminals by exchanging appropriate active (and reactive) power with the grid.

\subsection{GFMI Compliance Time-Domain Requirements}

To assess the VSBI characteristics of a GFMI, some SOs mandate and quantify response times as normative requirements. These active and reactive power (or current) response times are defined in time domain when the GFMI is subjected to a step change in voltage phase or magnitude at its terminals. However, these requirements have emerged only recently and remain at a relatively nascent stage of development. Broadly, they can be categorized into two groups: (i) predominantly qualitative requirements adopted by the National Energy System Operator (NESO, Great Britain) \cite{eso2021gc0137} and the Finnish Transmission System Operator (FINGRID) \cite{fingrid2023study}, and (ii) more detailed, time-domain quantitative specifications established by the Australian Energy Market Operator (AEMO) \cite{aemo2023voluntary}, the Electric Reliability Council of Texas (ERCOT) \cite{ERCOT2024AGS_ESR}, the Midcontinent Independent System Operator (MISO) \cite{miso2024gridforming} and the European Network of Transmission System Operators for Electricity (ENTSO-E) \cite{entsoe2024report}.

NESO defines active-phase-jump power (APJP) as the transient injection or absorption of active power by a GFMI due to changes in the phase angle between its IVS and the point of interconnection (POI) \cite{eso2021gc0137}. NESO requires APJP to be an inherent capability of a GFMI, with a natural response time of less than 5 milliseconds (ms). Similarly, FINGRID mandates a GFMI to provide phase jump (PJ) performance and system strength to resist voltage phase angle and magnitude changes. It mandates an initial response ``within a few ms'' and a ``full response'' within 10 ms \cite{fingrid2023study}.

AEMO has specified more detailed time-domain test setups and compliance criteria \cite{aemo2023voluntary}. AEMO recommends a single-GFMI infinite-bus (SGIB) test setup connected via variable impedance, with precise control over the voltage magnitude, frequency, and angle of the source \cite{aemo2023voluntary}. For the angle-step-change test, the short-circuit ratio (SCR) at the connection point is set to 3, and the system equivalent X/R ratio is set to 6. The initial dispatch of the GFMI is set to 0.5~p.u. The angle of the voltage source behind the equivalent grid impedance is decreased instantaneously by 10$^{\circ}$. To satisfy the compliance criterion, the instantaneous active power output of the GFMI should respond rapidly with a peak active power change of at least 0.2~p.u. for each 10$^{\circ}$ PJ \cite{aemo2023voluntary}. The response time to reach 90\% of the initial change in instantaneous active power should occur within 15~ms. Also, the active power should settle to the pre-disturbance level shortly after the PJ. 

Similar to AEMO \cite{aemo2023voluntary}, and for the same test setup and test sequence, ERCOT \cite{ERCOT2024AGS_ESR}, and MISO \cite{miso2024gridforming} also require at least 0.2~p.u. of peak active power for each 10$^{\circ}$ phase-angle step. ERCOT \cite{ERCOT2024AGS_ESR} mandates a maximum rise time of one cycle (16.66~ms for a 60~Hz system), while MISO \cite{miso2024gridforming} mandates a rise time of 15~ms. Similar to AEMO, MISO does not explicitly quantify a settling time and instead requires active power to settle to the pre-disturbance level shortly after the PJ. However, ERCOT \cite{ERCOT2024AGS_ESR} additionally mandates that the active power must be greater than or equal to the pre-disturbance level for at least three cycles (50~ms for a 60~Hz system). All three SOs also require that any oscillations settle and that any phase distortion dissipate over time. Similarly, ERCOT \cite{ERCOT2024AGS_ESR} also defines reactive power response requirements for small-signal voltage-magnitude jumps (MJ). In the same SGIB test setup, the impedance is set to zero, with the initial active power dispatch set to 1~p.u. and the initial reactive power dispatch set to 0~p.u. The magnitude of the voltage source behind the equivalent grid impedance is decreased instantaneously by 3\%. To satisfy the test criteria, the GFMI should exchange instantaneous reactive power with a peak of at least 0.03~p.u. for each 3\% step change in voltage magnitude \cite{ERCOT2024AGS_ESR}. The reactive power response time should be less than one cycle (16.66~ms for a 60~Hz system) and should not return to its pre-disturbance value within six cycles (100~ms for a 60~Hz system).

To ensure the generality of the requirement, ENTSO-E \cite{entsoe2024report} proposes defining the VSBI characteristics of GFMIs as a function of the effective reactance ($x_{{Eff}}$) between the IVS and the POI of the GFMI. It recommends a range of effective impedance values at 50~Hz to indicate the sensitivity of active (and reactive) current injection by a GFMI in response to variations in voltage angle and magnitude. ENTSO-E also recommends that SOs specify temporal parameters of dynamic performance to assess VSBI characteristics \cite{entsoe2024report}. Regarding dynamic performance for both voltage phase-angle and magnitude changes, it mandates a peak active current change between 50\% and 70\% of the RMS value calculated based on (\ref{eq:entsoe_ip_dut}). A maximum rise time of 10~ms is specified for active (and reactive) current/power variations. A decay time constant of 0.016~s, based on (\ref{eq:entsoe_tau}), is expected with a maximum error of 33\%. In cases where a steady-state value is required, a settling time of 60~ms is mandated, based on a 5\% tolerance band.

\begin{subequations}\label{eq:entsoe}
\begin{align}
i_{P,{DUT}} &\approx -\frac{u_{{Inv}}}{x_{{Eff}}} \sin(\delta) 
\label{eq:entsoe_ip_dut} \\[6pt]
\tau &= \frac{x_{{Eff}}}{k_f \omega_0} 
\label{eq:entsoe_tau}
\end{align}
\end{subequations}

where $u_{{Inv}}$ is the POI voltage, and $k_f$ is the droop constant of the GFMI. $\delta$ is the angle between the IVS and the POI of the GFMI, and $\omega_0$ is the grid frequency. Table~\ref{tab:gfm_requirements} summarizes the detailed, time-domain quantitative requirements from SOs from the second group.

\begin{table}[h]
  \caption{Summary of detailed time-domain requirements for GFMI VSBI characteristics specified by SOs.}
  \label{tab:gfm_requirements}
  \centering
  \renewcommand{\arraystretch}{1.15}
  \setlength{\tabcolsep}{2.2pt}
  \begin{tabularx}{\linewidth}{lX>{\raggedright\arraybackslash}p{1.3cm}X}
    \toprule
    \textbf{SO} & \textbf{Peak Value} & \textbf{Rise Time} & \textbf{Decay Time} \\
    \midrule
    AEMO & 
    $P \geq 0.2$ p.u./$10^{\circ}$ PJ &
    $\leq 15$ ms &
    -- \\
    ERCOT &
    $\begin{aligned}[t]
      P &\geq 0.2~\text{p.u.}/10^{\circ}\ \text{PJ} \\
      Q &\geq 0.03~\text{p.u.}/3\%\ \text{MJ}
    \end{aligned}$ &
    $\leq 16.6$ ms &
    $\begin{aligned}[t]
      \Delta P &\geq 0 \;\forall t \leq 50~\text{ms} \\
      \Delta Q &\geq 0 \;\forall t \leq 100~\text{ms}
    \end{aligned}$ \\
    MISO & 
    $P \geq 0.2$ p.u./$10^{\circ}$ PJ&
    $\leq 15$ ms &
    -- \\
    ENTSO-E &
    $I_{P}=50\text{--}70\%I_{\text{RMS}}$ (\ref{eq:entsoe_ip_dut})&
    $\leq 10$ ms &
    $t_{\text{decay}}=0.016$ s $\pm 33\%$ \\
    \bottomrule
  \end{tabularx}
\end{table}

\subsection{Issues with Measurement in the Time Domain}

\begin{figure}[htbp]
      \centering
      \includegraphics[width=0.9\linewidth]{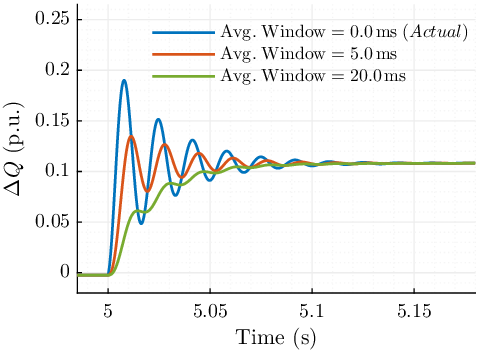}
      \caption{Simulation results demonstrating the effect of the measurement filter moving averaging window length on GFMI's change in reactive power response to a 5\% reduction in voltage magnitude at POI.}
      \label{fig:q_measurment_filter}
\end{figure}

The current time-domain approaches to compliance assessment by SOs \cite{aemo2023voluntary, ERCOT2024AGS_ESR, miso2024gridforming} of GFMIs have several major drawbacks. Since response times and, thus, current-time-domain requirements are at the sub-cycle or few-millisecond level, measurements of active and reactive power (or current) are highly sensitive to the measurement device's resolution, accuracy, and the length of the sliding window used in the moving-average filter. To smooth out noise, SOs recommend using a moving-average window with a maximum length specified in IEEE 2800 \cite{IEEE2800_2022} and IEC \cite{IEC61400_21_1_2019}, which is equal to one cycle (20 or 16.67~ms). In accordance with the reactive power response testing setup and the 3\% MJ recommended by ERCOT \cite{ERCOT2024AGS_ESR}, Fig.~\ref{fig:q_measurment_filter} illustrates the change in reactive power response of the generic National Lab of the Rockies (NLR, formerly NREL) GFMI model in PSCAD. It can be observed that the averaging window significantly affects the reactive-power response to an MJ at the POI, resulting in differences in rise time, peak value, and settling time.

\vspace{-0.2pt}
\subsection{GFMI Compliance Frequency-Domain Requirements}

Compliance assessment in frequency-domain tests can mitigate many of the challenges associated with time-domain tests. The short time frame in the time domain is easier to visualize in the frequency domain \cite{ESIG2025_GFM}. Since frequency-domain responses represent steady-state behavior at individual frequencies, they are largely independent of the measurement device's moving-average filter window. Furthermore, as discussed later, frequency-domain criteria for a GFMI can be defined independently of the testing setup (e.g., grid SCR and X/R ratio) as specified by SOs such as AEMO \cite{aemo2023voluntary}, ERCOT \cite{ERCOT2024AGS_ESR}, and MISO \cite{miso2024gridforming}.

Addressing compliance criteria in the frequency domain, \cite{testing_shahil_shah} presents a qualitative framework for testing the VSBI characteristics of GFMIs. It models the active (and reactive) power response of an ideal voltage source (IDVS) in the frequency domain. It qualitatively defines the expected ``IVS behind a reactor'' characteristics of a GFMI by expecting a nearly constant magnitude response and an approximately 180$^{\circ}$ phase response for the POI voltage phase-to-active-power ($P(s)/\theta(s)$) and POI voltage magnitude-to-reactive-power ($Q(s)/V(s)$) frequency scans over a frequency range from a few hertz (1--5~Hz) to a few tens of hertz (20--60~Hz). Similarly, \cite{Avdiaj2025VoltageSourceTechRxiv} qualitatively compares the essential VSBI characteristics of a Grid-FoLlowing Inverter (GFLI) and multiple GFMI control designs using the frequency-domain power-flow Jacobian. It recommends testing VSBI characteristics below 50~Hz and suggests that the magnitudes of the $P(s)/\theta(s)$ and $Q(s)/V(s)$ frequency responses within the frequency range of interest should be at least 0~dB (or 1~p.u./p.u.). However, these specifications remain largely qualitative, and to the best of the authors' knowledge, no work has yet quantitatively specified functional requirements or compliance criteria for the VSBI characteristics of a GFMI in the frequency domain.

Addressing these concerns and the lack of quantitative criteria in the frequency domain, this work proposes a frequency-domain compliance criterion for evaluating the VSBI characteristics of GFMIs. It maps the expected time-domain behavior to equivalent minimum $P(s)/\theta(s)$ and $Q(s)/V(s)$ Bode-plot characteristics within the frequency range of interest, thereby establishing the equivalence between time-domain and frequency-domain criteria. Compliance assessment based on the characteristics of the frequency-domain Jacobian elements, i.e., $P(s)/\theta(s)$ and $Q(s)/V(s)$, is shown to be straightforward and independent of the measurement device moving-average filter window and the testing setup. The proposed method is demonstrated by assessing the compliance of generic NLR GFMI models in PSCAD. Furthermore, case studies demonstrate the impact of GFMI compliance on the small-signal stability of the IEEE 39-bus system.


\section{Frequency-Domain Jacobian}

In the time domain, the relationship between the instantaneous active (and reactive) power output of an inverter-based resource (IBR) and the $d$- and $q$-axis components of the voltages and currents at its POI is given by \eqref{eq:pq_dq}:

\begin{subequations}\label{eq:pq_dq}
\begin{equation}
P_{IBR} = \frac{3}{2}\left(v_d i_d + v_q i_q\right)
\end{equation}
\begin{equation}
Q_{IBR} = \frac{3}{2}\left(-v_d i_q + v_q i_d\right)
\end{equation}
\end{subequations}

In the frequency domain, the $d$- and $q$-axis components of voltage and current at the POI of the IBR are related through the IBR admittance matrix $Y_{{IBR}}(s)$, as given in \eqref{eq:dq_admittance}:

\begin{equation}
-\begin{bmatrix}
I_d(s) \\
I_q(s)
\end{bmatrix}
=
\underbrace{
\begin{bmatrix}
Y_{dd}(s) & Y_{dq}(s) \\
Y_{qd}(s) & Y_{qq}(s)
\end{bmatrix}
}_{\mathbf{Y_{IBR}(s)}}
\begin{bmatrix}
V_d(s) \\
V_q(s)
\end{bmatrix}
\label{eq:dq_admittance}
\end{equation}

The Jacobian matrix in the frequency domain relates a IBR’s active ($P$) and reactive ($Q$) power responses to small perturbations in the voltage phase angle ($\theta$) and magnitude ($V$) at the POI \cite{testing_shahil_shah}. The Jacobian matrix $J_{{IBR}}(s)$ is defined as in \eqref{eq:jacobian_power}:

\begin{equation}
\begin{bmatrix}
P(s) \\
Q(s)
\end{bmatrix}
=
\underbrace{
\begin{bmatrix}
\dfrac{\partial P}{\partial V} 
& \dfrac{\partial P}{\partial \theta} \\[6pt]
\dfrac{\partial Q}{\partial V} 
& \dfrac{\partial Q}{\partial \theta}
\end{bmatrix}
}_{\mathbf{J_{IBR}}(s)}
\begin{bmatrix}
V(s) \\
\theta(s)
\end{bmatrix}
\label{eq:jacobian_power}
\end{equation}

As first introduced by Shah et al. in \cite{testing_shahil_shah}, the elements of the Jacobian matrix $J_{{IBR}}(s)$ are related to those of the admittance matrix $Y_{{IBR}}(s)$, as given in \eqref{eq:pq_vm_theta}.

\begin{equation}
\begin{bmatrix}
P(s) \\
Q(s)
\end{bmatrix}
=
\begin{bmatrix}
\dfrac{P_0}{V_0} - \dfrac{3}{2} V_0\, Y_{dd}
& - Q_0 - \dfrac{3}{2} V_0^2\, Y_{dq}
\\[6pt]
\dfrac{Q_0}{V_0} + \dfrac{3}{2} V_0\, Y_{qd}
& P_0 + \dfrac{3}{2} V_0^2\, Y_{qq}
\end{bmatrix}
\begin{bmatrix}
V(s) \\
\theta(s)
\end{bmatrix}
\label{eq:pq_vm_theta}
\end{equation}

where $P_0$, $Q_0$, and $V_0$ denote the pre-disturbance steady-state active power, reactive power, and voltage, respectively. The off-diagonal elements of the Jacobian matrix, i.e., $P(s)/\theta(s)$ and $Q(s)/V(s)$, are used to define the GFMI VSBI compliance criteria in the frequency domain.

\section{VSBI Characteristics in the Frequency Domain}

A GFMI is expected to exhibit VSBI characteristics in the sub-transient to transient time frame. The realization and testing of this requirement is near-instantaneous dispatch of active and/or reactive power following a grid disturbance. When a PJ (or MJ) occurs at the POI of a GFMI, the active (or reactive) power response is expected to change rapidly. The speed and magnitude of the GFMI response qualitatively characterize its VSBI strength. Thus, a GFMI exhibiting a rapid and significant active (and reactive) power response to a PJ (or MJ) is considered to exhibit strong VSBI characteristics. Conversely, a GFMI exhibiting a slow and/or low-magnitude response is considered to exhibit weak VSBI characteristics.

For an ideal voltage source behind an impedance, the response settles to a new value after some time (referred to as the quasi-steady state). However, for a GFMI, the response in quasi-steady state can also be expected to return to its original value over time due to droop control. Thus, the expected VSBI characteristics of a GFMI can be divided into (a) quick response dynamics dominated by the faster inner control (IC) loops and (b) subsequent decay dynamics depending on the design of the slower outer synchronization (OS) loops. As discussed next, in the frequency domain, the quick-response dynamics, dominated by the faster IC loops, follow standard second-order low-pass filter (LPF) characteristics and shape the $P(s)/\theta(s)$ and $Q(s)/V(s)$ characteristics in the higher-frequency range. Similarly, the slower decay dynamics, governed by the OS loops, follow standard second-order high-pass filter (HPF) characteristics and shape the $P(s)/\theta(s)$ and $Q(s)/V(s)$ characteristics in the lower-frequency range.

Immediately after a disturbance, the GFMI is expected to exchange active (or reactive) power with the grid, similar to an ideal voltage source behind an impedance. Therefore, in the frequency domain, a GFMI is expected to exhibit similar $P(s)/\theta(s)$ and $Q(s)/V(s)$ characteristics. 
The $dq$-domain admittance $Y_{{Id}}(s)$ of an ideal voltage source behind an impedance ($R + j\omega_0 L$) is given in \eqref{eq:Yi_matrix}.

\begin{equation}
\mathbf{Y}_{Id}(s)
=
\frac{1}{(R + sL)^2 + (\omega_0 L)^2}
\begin{bmatrix}
R + sL & \omega_0 L \\
-\omega_0 L & R + sL
\end{bmatrix}
\label{eq:Yi_matrix}
\end{equation}

The off-diagonal terms of the Jacobian matrix, representing the active (and reactive) power response to a PJ (and MJ), are derived from \eqref{eq:Yi_matrix} and \eqref{eq:pq_vm_theta} and are given in \eqref{eq:pq_partials}.

\begin{subequations}\label{eq:pq_partials}
\begin{equation}
\left.\frac{P(s)}{\theta(s)}\right|_{V(s)=0}
=
- Q_0
-
\frac{3}{2} V_0^2 \,
\frac{\omega_0 L}{(R + sL)^2 + (\omega_0 L)^2}
\end{equation}

\begin{equation}
\left.\frac{Q(s)}{V(s)}\right|_{\theta(s)=0}
=
\frac{Q_0}{V_0}
-
\frac{3}{2} V_0 \,
\frac{\omega_0 L}{(R + sL)^2 + (\omega_0 L)^2}
\end{equation}
\end{subequations}

From \eqref{eq:pq_partials}, it can be observed that around an operating point, both the $P(s)/\theta(s)$ and $Q(s)/V(s)$ transfer functions of an ideal voltage source behind an impedance exhibit standard second-order LPF characteristics, with poles at $s = -\tfrac{R}{L} \pm j\omega_{0}$ and a damped natural frequency of $\omega_d = \omega_0$. Thus, for a critically compliant GFMI (i.e., just on the threshold of violating compliance criteria) with the specified maximum rise-time and minimum peak-value constraints, similar standard second-order LPF characteristics are expected in $P(s)/\theta(s)$ (and $Q(s)/V(s)$). As discussed in the next section, these boundary LPF characteristics are directly determined by the specified maximum rise-time and minimum peak-value time-domain criteria of the active (and reactive) power responses.

\begin{figure}[htbp]
      \centering
      \includegraphics[width=0.9\linewidth]{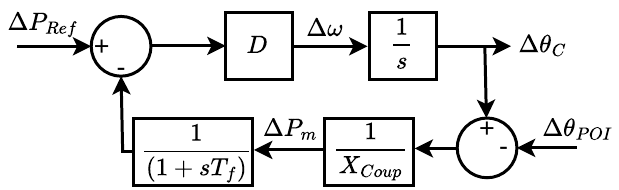}
      \caption{Block diagram representing the small-signal dynamics for droop-based outer synchronization (OS) loop of a GFMI.}
      \label{fig:droop_diag}
\end{figure}

Similarly, if the GFMI response is expected to return to its original value in the quasi-steady state, a minimum decay time is specified. This decay dynamic in a GFMI depends on the design of the slower OS loop, specifically the active (and/or reactive) power droop control loops. A typical GFMI design ensures sufficient bandwidth separation between the OS and IC loops. For the generic NLR GFMI model in PSCAD \cite{kenyon2021open}, assuming the IC loops are fast, the small-signal dynamics of the most common droop-based OS loop is illustrated in Fig.~\ref{fig:droop_diag}. 

Where $P_{{Ref}}$, $P_m$, $\omega$, $\theta_C$, $\theta_{{POI}}$, and $T_f$ denote the active power reference, active power output, GFMI frequency, voltage angle at the filter capacitor, voltage angle at the POI, and active power measurement filter time constant, respectively. $\Delta(\cdot)$ denotes the small-signal representation of the respective variables. The small-signal active power flow across the coupling reactance ($X_{{Coup}}$) is given in \eqref{eq:power_margin}.

\begin{equation}
\Delta P_m = \frac{\Delta \theta_C - \Delta \theta_{{POI}}}{X_{{Coup}}}
\label{eq:power_margin}
\end{equation}

The small-signal GFMI frequency as a function of the droop loop is given in \eqref{eq:domega_dpref}.

\begin{equation}
\Delta \omega
=
D\left(
\Delta P_{{Ref}}
-
\frac{1}{1 + s T_f}\, \Delta P_m
\right)
\label{eq:domega_dpref}
\end{equation}

The small-signal voltage angle at the filter capacitor is given in \eqref{eq:delta_from_omega}.

\begin{equation}
\Delta \theta_C = \frac{1}{s}\,\Delta \omega
\label{eq:delta_from_omega}
\end{equation}

For a constant active power reference, combining \eqref{eq:domega_dpref} and \eqref{eq:delta_from_omega} yields the small-signal voltage angle at the filter capacitor, as given in \eqref{eq:delta_combined}.

\begin{equation}
\Delta \theta_C 
= 
- \frac{D}{s(1 + s T_f)}\, \Delta P_m
\label{eq:delta_combined}
\end{equation}

Substituting $\Delta \theta_C$ from \eqref{eq:delta_combined} into \eqref{eq:power_margin} yields \eqref{eq:pm_substituted}.

\begin{equation}
\Delta P_m
=
\frac{
-\dfrac{D}{s(1+sT_f)} \Delta P_m
-
\Delta \theta_{{POI}}
}
{X_{{Coup}}}
\label{eq:pm_substituted}
\end{equation}

Simplifying \eqref{eq:pm_substituted} yields the closed-loop transfer function from $\Delta \theta_{{POI}}$ to $\Delta P_m$, as given in \eqref{eq:pm_over_deltaPOI_poly}.


\begin{equation}
\frac{\Delta P_m}{\Delta \theta_{{POI}}}
(\text{or simply}\;
\frac{P(s)}{\theta(s)})
=
-
\frac{T_f s^2 + s}
{
X_{{Coup}} T_f s^2 + X_{{Coup}} s + D
}
\label{eq:pm_over_deltaPOI_poly}
\end{equation}

This $P(s)/\theta(s)$ transfer function exhibits HPF characteristics, with a DC gain ($s \rightarrow 0$) that rolls off to zero and a constant high-frequency gain ($s \rightarrow \infty$) of $1/X_{{Coup}}$. These characteristics are consistent with those of a standard HPF, particularly above the cutoff frequency of the active power filter, $1/T_f$, which is typically in the range of a few hertz for a GFMI \cite{epri2024gridforming}. Similarly, the $Q(s)/V(s)$ characteristics can be derived when the GFMI operates in reactive power dispatch mode with a reactive power droop loop. Thus, for a GFMI that is critically compliant with the specified minimum decay-time and minimum peak-value constraints, similar standard second-order HPF characteristics are expected in $P(s)/\theta(s)$ (and $Q(s)/V(s)$). As discussed in the next section, these boundary HPF characteristics are directly determined by the specified minimum decay-time and minimum peak-value time-domain criteria of the active (and reactive) power responses.

\section{Frequency-Domain Compliance Criteria}

\subsection{Design of Frequency-Domain Compliance Criteria}

As summarized in Table~\ref{tab:gfm_requirements}, for the compliance assessment of GFMIs, SOs mandate three key characteristics for the active and reactive power responses. As discussed in Section~II, the maximum rise time and minimum peak value primarily depend on the IC loops, thereby constraining the second-order LPF characteristics at higher frequencies. Similarly, the decay time depends on the OS loop and thus constrains the second-order HPF characteristics at lower frequencies. Accordingly, the frequency-domain compliance criteria can be defined through two boundary conditions: (a) second-order LPF characteristics in the higher-frequency range and (b) second-order HPF characteristics in the lower-frequency range.

The first boundary $P(s)/\theta(s)$ (and $Q(s)/V(s)$) characteristics are defined by mapping them to the expected rise time and peak value of the active (and reactive) power response, using the standard form of a second-order LPF given in \eqref{standard_LPF}.

\begin{equation}
LP(s) = K_{{LPF}} \cdot 
\frac{\omega_{n,{LPF}}^{2}}
{s^{2} + 2\zeta_{{LPF}} \omega_{n,{LPF}} s + \omega_{n,{LPF}}^{2}}
\label{standard_LPF}
\end{equation}

\begin{figure}[htbp]
    \centering
    
        \begin{subfigure}{0.485\linewidth}
        \centering
        \includegraphics[width=\linewidth]{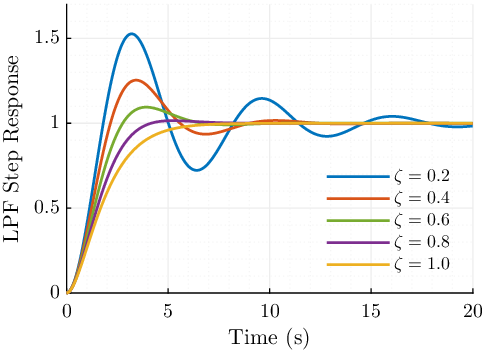}
        \caption{}
        \label{fig:step_response}
    \end{subfigure}
    \hfill
    \begin{subfigure}{0.485\linewidth}
        \centering
        \includegraphics[width=\linewidth]{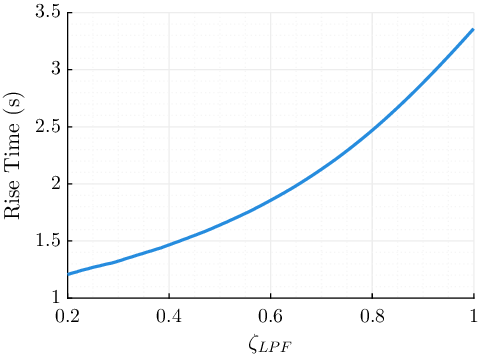}
        \caption{}
        \label{fig:rise_time}
    \end{subfigure}
    \hfill
    \caption{Effect of varying damping ratio ($\zeta_{LPF}$) on a second-order LPF with $\omega_{n,LPF}=1$: (a). step response, and (b). rise time.}
    \label{fig:zeta_plots}
\end{figure}

This equation has three tunable parameters: $\zeta_{{LPF}}$, $\omega_{n,{LPF}}$, and $K_{{LPF}}$, which must be determined for the boundary LPF case. For a GFMI, due to a high X/R ratio, the $P(s)/\theta(s)$ (and $Q(s)/V(s)$) responses are generally under-damped ($\zeta_{{LPF}} \leq 1$) \cite{aemo2023voluntary}. For a standard LPF system, the variation in the step response and rise time with varying $\zeta_{{LPF}}$ (with $\omega_{n,{LPF}} = 1$) is shown in Figs.~\ref{fig:step_response} and \ref{fig:rise_time}, respectively. It can be observed that the critically damped case ($\zeta_{{LPF}} = 1$) results in the slowest rise time and the lowest peak (equal to the steady-state value) and is thus the most conservative scenario. However, SOs may choose $\zeta_{{LPF}}$ to represent a more realistic case rather than the most conservative one.

 For a standard second-order underdamped LPF, the widely accepted empirical formula to calculate the rise time ($t_{r,{LPF}}$) is given in \eqref{tr_lpf} \cite{nise2015control}. Thus, for the LPF boundary case with $\zeta_{{LPF}} = 1$, the natural frequency $\omega_{n,{LPF}}$ can be calculated from the specified maximum rise time using \eqref{tr_lpf}. Similarly, for a standard LPF, the critically damped boundary case ($\zeta_{{LPF}} = 1$) exhibits no overshoot. Hence, the DC gain $K_{{LPF}}$ is equal to the minimum expected peak value. Substituting these parameters into \eqref{standard_LPF} yields the required second-order LPF boundary $P(s)/\theta(s)$ (and $Q(s)/V(s)$) characteristics for a GFMI.

\begin{equation}
t_{r,{LPF}} =
\frac{
1.76\,\zeta_{{LPF}}^{3}
- 0.417\,\zeta_{{LPF}}^{2}
+ 1.039\,\zeta_{{LPF}}
+ 1
}{
\omega_{n,{LPF}}
}
\label{tr_lpf}
\end{equation}

Similarly, the second boundary $P(s)/\theta(s)$ (and $Q(s)/V(s)$) characteristics are defined by mapping them to the expected decay time and peak value of the active (and reactive) power response, using the standard form of a second-order HPF given in \eqref{standard_HPF}.

\begin{equation}
HP(s) = K_{{HPF}} \cdot 
\frac{s^{2}}
{s^{2} + 2\zeta_{{HPF}} \omega_{n,{HPF}} s + \omega_{n,{HPF}}^{2}}
\label{standard_HPF}
\end{equation}

The standard HPF equation has three tunable parameters: $\zeta_{{HPF}}$, $\omega_{n,{HPF}}$, and $K_{{HPF}}$, which must be determined for the boundary HPF case. Similar to the LPF characteristics, the critically damped HPF case ($\zeta_{{HPF}} = 1$) represents the most conservative scenario due to the quickest decay from its peak value. Given sufficient bandwidth separation between the quick response dynamics and the subsequent decay dynamics of a GFMI, the time-domain response settles to its steady-state (peak) value before the decay dynamics begin. Thus, for the critically damped HPF boundary case with $\zeta_{{HPF}} = 1$ and no undershoot, the constant $K_{{HPF}}$ is equal to the minimum expected peak value.

To determine the natural frequency $\omega_{n,{HPF}}$ for the HPF boundary case, an empirical decay-time ($t_{d,{HPF}}$) relationship is derived for a standard second-order underdamped HPF. This approach is analogous to the standard empirical rise-time expression in \eqref{tr_lpf} for a standard second-order underdamped LPF. The empirical decay-time relationship is obtained using cubic polynomial fitting of $\zeta_{{HPF}}$ values for $\omega_{n,{HPF}} = 1$ and is given in \eqref{td_hpf}.

\begin{equation}
t_{d,{HPF}} =
\frac{
-0.09\,\zeta_{{HPF}}^{3}
+ 0.42\,\zeta_{{HPF}}^{2}
- 0.88\,\zeta_{{HPF}}
+ 1.56
}{
\omega_{n,{HPF}}
}
\label{td_hpf}
\end{equation}

\begin{figure}[htbp]
      \centering
      \includegraphics[width=0.9\linewidth]{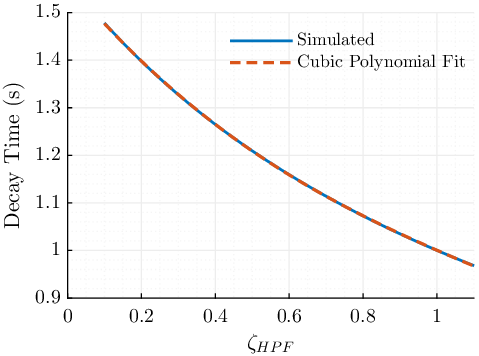}
      \caption{Fitting accuracy for the simulated decay time from the step response vs cubic polynomial fit (poly3) for standard second-order HPF in \eqref{standard_HPF} with varying $\zeta_{HPF}$ \& $\omega_{n,HPF}=1$.}
      \label{fig:t_decay_fit}
\end{figure}

The fitting accuracy for the simulated decay time versus the cubic polynomial (poly3) fit of a standard HPF as a function of $\zeta_{{HPF}}$ is illustrated in Fig.~\ref{fig:t_decay_fit}. The fit exhibits excellent accuracy, with goodness-of-fit metrics given by the coefficient of determination $R^2 = 0.99$, adjusted $R^2 = 0.99$, and root-mean-square error ${RMSE} = 0.92 \times 10^{-4}$.

Thus, for the HPF boundary case with $\zeta_{{HPF}} = 1$, the natural frequency $\omega_{n,{HPF}}$ can be calculated from the specified minimum decay time using \eqref{td_hpf}. Substituting the obtained $\zeta_{{HPF}}$, $\omega_{n,{HPF}}$, and $K_{{HPF}}$ into \eqref{standard_HPF} yields the required second-order HPF boundary $P(s)/\theta(s)$ (and $Q(s)/V(s)$) characteristics for a GFMI.

Given that \eqref{eq:pm_over_deltaPOI_poly} and \eqref{eq:pq_partials} are minimum-phase transfer functions, the Bode magnitude and phase plots are uniquely related and thus both the LPF and HPF compliance criteria can be defined solely from the magnitude plot \cite{OppenheimSchafer2009}. The speed of response (rise time) of the second-order LPF system is primarily determined by its $-3$~dB bandwidth frequency $f_{{BW,LPF}}$, i.e., the maximum frequency up to which the GFMI should respond effectively to fast transients without suppressing high-frequency disturbances. Similarly, the speed of response (decay time) of the second-order HPF system is primarily determined by its $-3$~dB bandwidth frequency $f_{{BW,HPF}}$, i.e., the minimum frequency down to which the GFMI should respond effectively to slow transients without suppressing low-frequency disturbances. Thus, $f_{{BW,LPF}}$ and $f_{{BW,HPF}}$ define the upper and lower limits, respectively, of the compliance frequency range for the minimum Bode magnitude. The LPF boundary condition, dominated by the faster IC loops, takes precedence at higher frequencies, whereas the HPF boundary condition, dominated by the slower OS loops, takes precedence at lower frequencies. Therefore, for a GFMI to be compliant, its Bode magnitude should be greater than or equal to the defined $HP(s)$ from $f_{{BW,HPF}}$ to $f_{{Int}}$ and greater than or equal to the defined $LP(s)$ from $f_{{Int}}$ to $f_{{BW,LPF}}$, where $f_{{Int}}$ denotes the intersection frequency of the LPF and HPF boundary conditions. The frequency-domain compliance criteria corresponding to AEMO, MISO, and ERCOT are derived and illustrated in the next section.

\subsection{Time-to-Frequency-Domain Conversion of VSBI Compliance Criteria from AEMO, MISO, and ERCOT}

This subsection demonstrates the conversion of the VSBI time-domain compliance criteria specified by AEMO, MISO, and ERCOT into equivalent frequency-domain minimum $P(s)/\theta(s)$ and $Q(s)/V(s)$ Bode-magnitude criteria. As discussed in Section~III, AEMO's rise-time and peak-value criteria for the active power response to a voltage-angle step change are converted into an equivalent frequency-domain $P(s)/\theta(s)$ criterion. For a rise time of 15~ms (as given in Table~\ref{tab:gfm_requirements},) and the conservative choice of $\zeta_{{LPF}}^{{AEMO}} = 1$, the natural frequency $f_{n,{LPF}}^{{AEMO}} = \omega_{n,{LPF}}^{{AEMO}}/2\pi$, computed from \eqref{tr_lpf}, is 33.88~Hz, and the computed $-3$~dB bandwidth frequency $f_{{BW,LPF}}^{{AEMO}}$ for standard $LP(s)$ in\eqref{standard_LPF} is 23.09~Hz. The gain $K_{{LPF}}^{{AEMO}} = 0.2/0.174$ is calculated based on the expected active power peak value of 0.2~p.u. per 10$^\circ$ PJ. The resulting boundary LPF characteristic for AEMO is given in \eqref{eq:LP_AEMO}.

\begin{equation}
\label{eq:LP_AEMO}
LP_{AEMO}(s) =
\frac{5.825 \times 10^{4}}{s^{2} + 450.9\,s + 5.084 \times 10^{4}}
\end{equation}

\begin{figure}[htbp]
    \centering
    \includegraphics[width=0.9\linewidth]{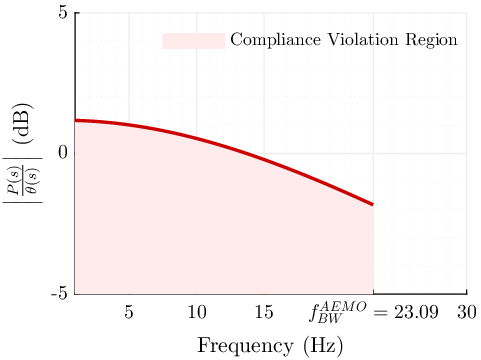}
    \caption{Minimum $P(s)/\theta(s)$ magnitude Bode compliance envelope derived from AEMO time-domain criteria.}
    \label{fig:AEMO_min_bode}
\end{figure}

The minimum $P(s)/\theta(s)$ magnitude Bode-plot compliance boundary for AEMO is shown as the red shaded region in Fig.~\ref{fig:AEMO_min_bode}. This envelope provides a clear and intuitive visualization of the proposed frequency-domain compliance criterion for AEMO.

\begin{figure}[htbp]
    \centering
    \includegraphics[width=0.9\linewidth]{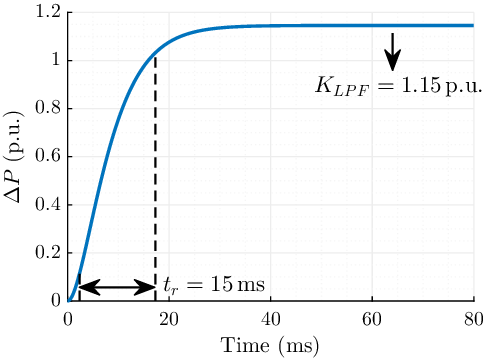}
    \caption{Active-power step response of a critically compliant GFMI validating equivalence between AEMO time-domain and frequency-domain criteria in Fig.~\ref{fig:AEMO_min_bode}.}
    \label{fig:AEMO_min_bode_step}
\end{figure}

Figure~\ref{fig:AEMO_min_bode_step} illustrates the step response of a GFMI that is critically compliant in the frequency domain, i.e., a GFMI with $|P(s)/\theta(s)|$ exactly equal to the minimum $P(s)/\theta(s)$ magnitude Bode-plot compliance boundary shown in Fig.~\ref{fig:AEMO_min_bode}. It can be verified from Fig.~\ref{fig:AEMO_min_bode_step} that the critically compliant GFMI has a rise time of 15~ms and a peak (and steady-state) value of 1.15~p.u., which exactly matches the AEMO time-domain compliance criteria. Since MISO specifies identical time-domain criteria as AEMO, it also results in the same minimum-magnitude Bode-plot compliance boundary characteristics shown in Fig.~\ref{fig:AEMO_min_bode}.

Similarly, ERCOT's rise-time and peak-value criteria for the reactive power response to a voltage-magnitude step change is converted into an equivalent frequency-domain $Q(s)/V(s)$ LPF criterion. As summarized in Table~\ref{tab:gfm_requirements}, for a rise time of 16.67~ms and the conservative choice of $\zeta_{{LPF}}^{{ERCOT}} = 1$, the natural frequency $f_{n,{LPF}}^{{ERCOT}} = \omega_{n,{LPF}}^{{ERCOT}}/2\pi$, computed from \eqref{tr_lpf}, is 32.3~Hz, and the $-3$~dB bandwidth frequency $f_{{BW,LPF}}^{{ERCOT}}$ is 20.8~Hz. The gain $K_{{LPF}}^{{ERCOT}} = 0.03/0.03$ is calculated based on the expected reactive power peak value of 0.03~p.u. per 3\% MJ. The resulting boundary LPF characteristic for ERCOT is given in \eqref{eq:LP_ERCOT}.

\begin{equation}
\label{eq:LP_ERCOT}
LP_{ERCOT}(s) =
\frac{4.118 \times 10^{4}}{s^{2} + 405.8\,s + 4.118 \times 10^{4}}
\end{equation}

The $Q(s)/V(s)$ HPF criterion for ERCOT is derived from the specified decay-time and peak-value criteria for the reactive power response to a voltage-magnitude step change. For a decay time of 100~ms and the conservative choice of $\zeta_{{HPF}}^{{ERCOT}} = 1$, the natural frequency $f_{n,{HPF}}^{{ERCOT}} = \omega_{n,{HPF}}^{{ERCOT}}/2\pi$, computed from \eqref{td_hpf}, is 1.59~Hz, and the computed $-3$~dB bandwidth frequency $f_{{BW,HPF}}^{{ERCOT}}$ for standard $HP(s)$ in\eqref{standard_HPF} is 2.47~Hz. The resulting boundary HPF characteristic for ERCOT is given in \eqref{eq:HP_ERCOT}.

\begin{equation}
\label{eq:HP_ERCOT}
HP_{ERCOT}(s) =
\frac{s^{2}}{s^{2} + 20\,s + 100}
\end{equation}

\begin{figure}[htbp]
    \centering
    \includegraphics[width=0.9\linewidth]{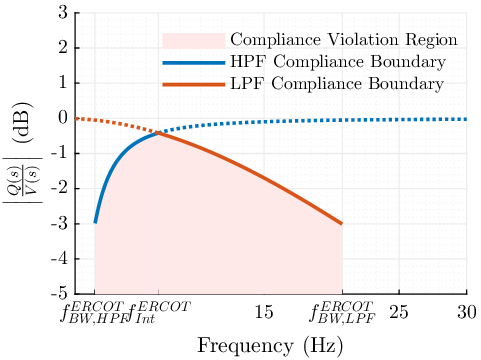}
    \caption{Minimum $Q(s)/V(s)$ magnitude Bode compliance envelope derived from ERCOT reactive-power criteria.}
    \label{fig:ERCOT_min_bode}
\end{figure}

The minimum $Q(s)/V(s)$ magnitude Bode-plot compliance boundary for ERCOT is shown as the red shaded region in Fig.~\ref{fig:ERCOT_min_bode}. This envelope is defined by the combination of the LPF and HPF compliance boundary conditions. The intersection frequency of the LPF and HPF boundary conditions, $f_{{Int}}^{{ERCOT}}$, is 7.17~Hz. Thus, from $f_{{BW,HPF}}^{{ERCOT}} = 2.47$~Hz to $f_{{Int}}^{{ERCOT}} = 7.17$~Hz, the compliance boundary is governed by the HPF characteristics, whereas from $f_{{Int}}^{{ERCOT}} = 7.17$~Hz to $f_{{BW,LPF}}^{{ERCOT}} = 20.8$~Hz, it is governed by the LPF characteristics. The envelope shown in Fig.~\ref{fig:ERCOT_min_bode} provides a clear and intuitive visualization of the proposed frequency-domain compliance criterion for ERCOT. The $P(s)/\theta(s)$ frequency-domain compliance criterion for ERCOT can be derived in a similar manner. The validation and application of the compliance criteria are demonstrated in the next section.

\section{Case Studies}

This section evaluates the compliance of generic NLR GFMI models in PSCAD using the proposed frequency-domain compliance criteria. The impact of compliant and non-compliant GFMIs on oscillations and small-signal stability is subsequently assessed using the IEEE 39-bus (10-generator) bulk-power system.

\subsection{Compliance of Generic NLR GFMI models in PSCAD}

\begin{figure}[htbp]
      \centering
      \includegraphics[width=0.9\linewidth]{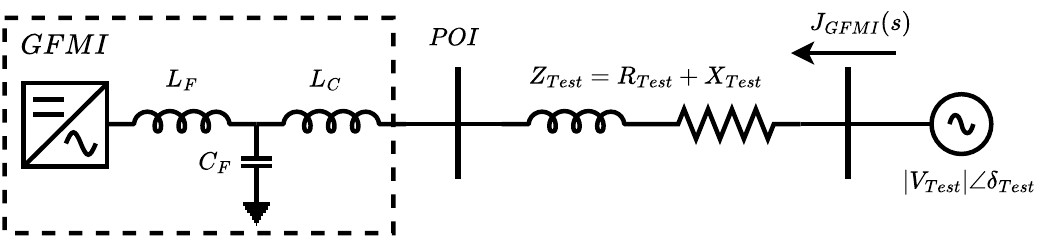}
      \caption{GFMI frequency-domain Jacobian scan test setup.}
      \label{fig:test_setup}
\end{figure}

To evaluate GFMI compliance in the frequency domain, Fig.~\ref{fig:test_setup} illustrates the test setup used to obtain the $P(s)/\theta(s)$ and $Q(s)/V(s)$ elements of the frequency-domain Jacobian, $J_{{GFMI}}(s)$. The test voltage $|V_{{Test}}| \angle \delta_{{Test}}$ is perturbed to observe the active and reactive power responses and thereby extract the Jacobian elements. It is important to note that the time-domain specifications for active and reactive power are defined for a specific test setup (as described in Section~II); therefore, the frequency-domain Jacobian elements must be obtained by accounting for the test-setup impedance, $Z_{{Test}}$. However, the proposed frequency-domain compliance criteria remain independent of the test setup. If the time-domain specifications are defined at the POI of the GFMI, compliance (via the proposed frequency-domain criteria) can be assessed using the frequency-domain Jacobian evaluated at the POI of the GFMI.

\begin{figure}[htbp]
      \centering
      \includegraphics[width=0.9\linewidth]{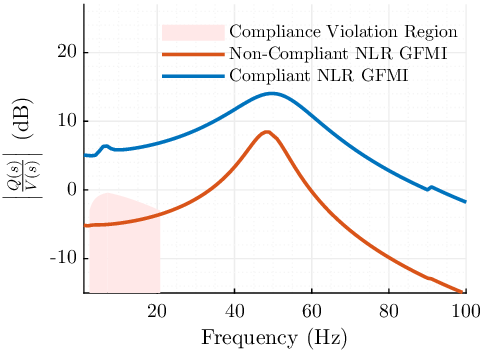}
      \caption{ERCOT $Q(s)/V(s)$ compliance assessment of two NLR GFMI designs; one satisfies and one violates the minimum magnitude criterion.}
      \label{fig:NLR_bode}
\end{figure}

We now evaluate the compliance of the generic GFMI model developed by NLR in PSCAD with respect to ERCOT's reactive power response criteria. As described in Section~II, $Z_{{Test}}$ is set to zero, and the GFMIs are dispatched at P = 1 p.u. and Q = 0 p.u.. The control parameters and hardware impedances of this model, described in \cite{kenyon2021open}, are adjusted within typical bandwidths and ranges \cite{chatterjee2025unifi} to obtain two GFMI designs exhibiting weak and strong VSBI characteristics. However, for compliance assessment, these models are treated as black-box representations. ERCOT's reactive power response compliance assessment of the NLR GFMI is illustrated in Fig.~\ref{fig:NLR_bode}. It can be observed that the GFMI design (in orange) has a $Q(s)/V(s)$ Bode magnitude below the minimum $Q(s)/V(s)$ compliance threshold and thus fails the compliance test.

\begin{figure}[htbp]
      \centering
      \includegraphics[width=0.9\linewidth]{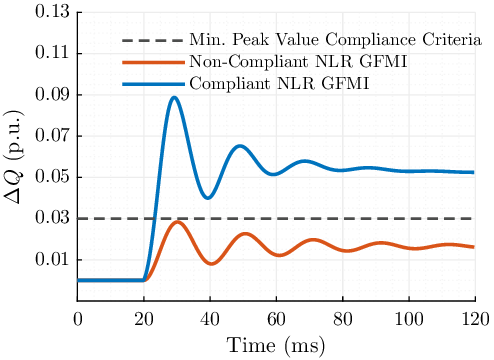}
      \caption{Change in reactive power response of compliant and non-compliant NLR GFMIs in PSCAD for a 3\% drop in $|V_{Test}|$, confirming frequency-domain assessment.}
      \label{fig:NLR_step}
\end{figure}

To verify the validity of the proposed frequency-domain compliance criteria, the test-setup voltage magnitude ($|V_{{Test}}|$) is stepped from 1~p.u. to 0.97~p.u.. The resulting change in reactive power responses of the GFMIs is illustrated in Fig.~\ref{fig:NLR_step}. With a peak reactive power change of 0.028~p.u., the non-compliant NLR GFMI design (in orange) fails to meet the specified peak-value criterion as summarized in Table~\ref{tab:gfm_requirements}. The compliant NLR GFMI design (in blue) exhibits a rise time of 3.6~ms and a peak value of 0.08~p.u., demonstrating compliance with the LPF characteristics. Furthermore, the reactive power response of the compliant NLR GFMI design does not drop below zero within 100~ms, thereby satisfying the HPF characteristics. Similarly, the GFMI can be evaluated for compliance with the active power response criterion, which is not illustrated here due to space constraints. These results demonstrate that the proposed method provides a clear and intuitive visualization of VSBI compliance in the frequency domain.

\subsection{Impact on GFMI Compliance on Small-Signal Stability}

\begin{figure}[htbp]
    \centering
    \includegraphics[width=1\linewidth]{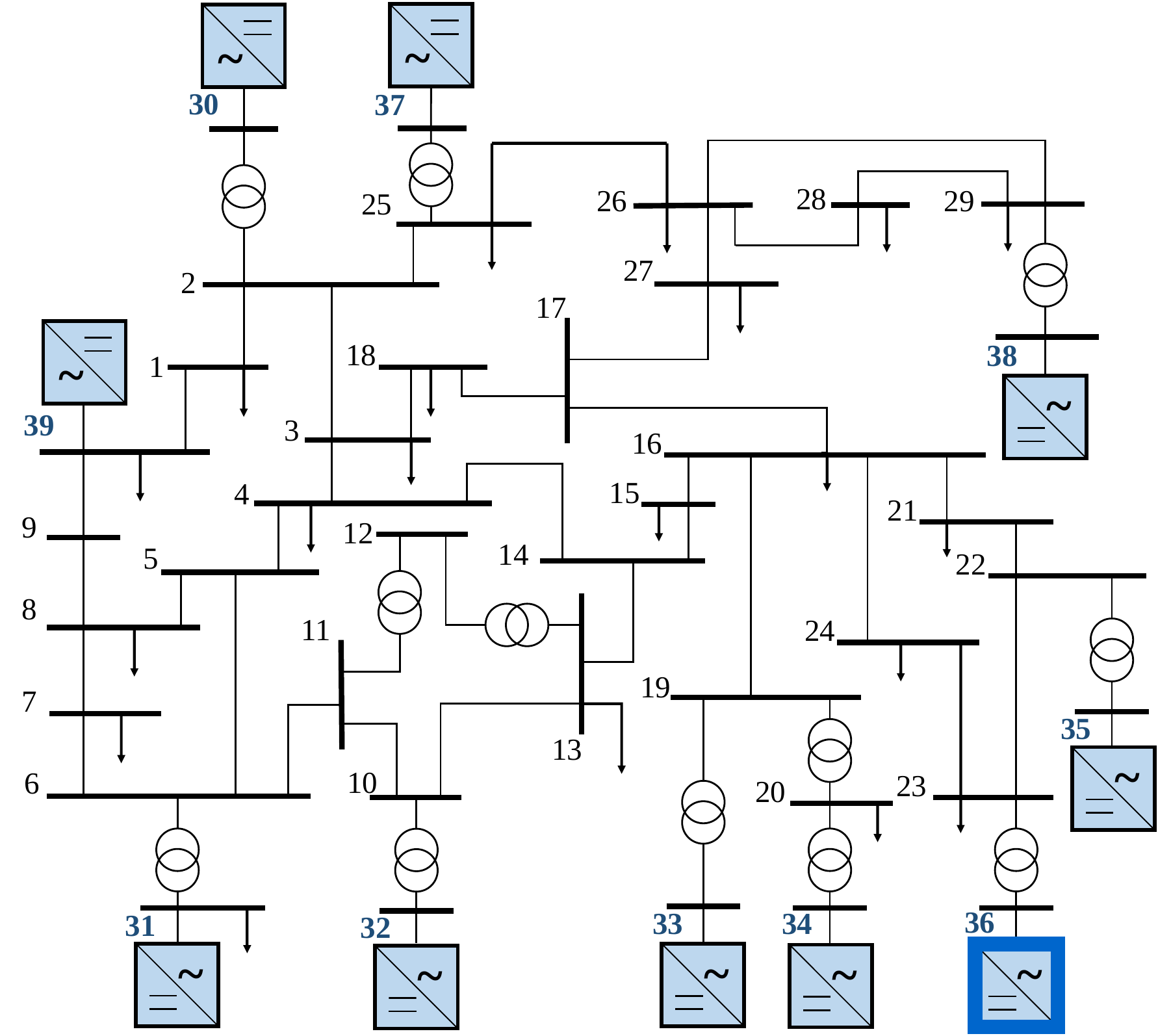}
    \caption{A modified IEEE 39-bus test system \cite{ref39bus}, with all machines replaced by IBRs and a GFMI at Bus 36. See Table \ref{tab:gfl_gfm_scenarios} for the type and location of all IBRs.}
    \label{fig:39Bus_c}
\end{figure}

In this subsection, the impact of compliant and non-compliant GFMIs on small-signal stability in bulk power systems is analyzed. The impact of VSBI compliance of GFMIs on IBR-induced weak grid oscillations is investigated using a modified IEEE 39-bus test system \cite{ref39bus}. The IEEE 39-bus system is modified by replacing all 10 synchronous generators with IBRs, as shown in Fig.~\ref{fig:39Bus_c}.

It is well understood that GFLIs require a sufficiently strong grid to operate stably and can induce phase-locked loop (PLL)-driven oscillations and small-signal stability issues in weak grids \cite{esig2024oscillations}. In contrast, a GFMI, owing to its voltage-source or VSBI characteristics, can enhance grid strength and help damp weak-grid oscillations. Thus, to highlight the impact of GFMI VSBI compliance on small-signal stability, the 39-bus system is modified to include only one GFMI at Bus~36, while all other generators are replaced by GFLIs. The IBR configurations and dispatch scenario for buses 30--39 are summarized in Table~\ref{tab:gfl_gfm_scenarios}. The GFLI and GFMI models used in this case study are adopted from \cite{emt-rms2}. To clarify, fully detailed (white-box) models are often unavailable to SOs and are not required for compliance testing with the proposed method. For compliance purposes, these models are treated as black-box representations. The detailed white-box models are used only for root-cause analysis of poorly damped oscillations and for identifying vulnerable sections of the grid.

\begin{table}[htbp]
\centering
\caption{IBR configurations and dispatch scenario at IBR buses 30--39 in the modified IEEE 39-bus system.}
\label{tab:gfl_gfm_scenarios}
\renewcommand{\arraystretch}{1.15}
\setlength{\tabcolsep}{3pt}
\footnotesize
\resizebox{\columnwidth}{!}{%
\begin{tabular}{|l|c|c|c|c|c|c|c|c|c|c|}
\hline
\textbf{Bus}   & \textbf{30}  & \textbf{31}  & \textbf{32}  & \textbf{33 } & \textbf{34}  & \textbf{35}  & \textbf{36}  & \textbf{37}  & \textbf{38} & \textbf{39}  \\ \hline
\textbf{Type}  & GFLI & GFLI & GFLI & GFLI & GFLI & GFLI & \textit{GFMI} & GFLI & GFLI & GFLI \\ \hline
\textbf{P (pu)} & 2.75 & 5.21 & 6.50 & 6.32 & 5.08 & 6.50 & 5.60 & 5.40 & 8.30 & 9.77 \\ \hline
\textbf{Q (pu)} & 1.69 & 1.42 & 1.47 & 0.49 & 1.38 & 2.32 & 1.70 & 0.16 & 0.48 & 3.51 \\ \hline
\end{tabular}%
}
\end{table}

\begin{figure}[htbp]
    \centering
    \includegraphics[width=0.9\linewidth]{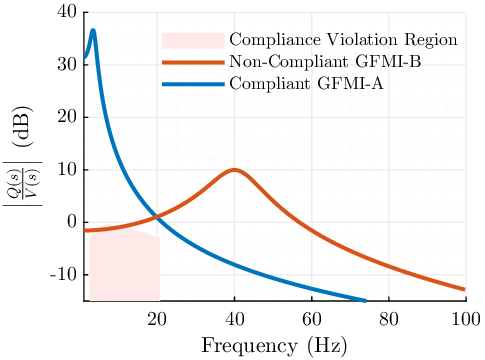}
    \caption{ERCOT's $Q(s)/V(s)$ frequency domain compliance assessment for GFMI-A and GFMI-B for analysing small-signal stability issues in the modified IEEE-39 Bus system.}
    \label{fig:matlab_compliance}
\end{figure}

To assess the impact of GFMI compliance on small-signal stability, the control parameters and hardware impedances of the GFMI at Bus~36 are modified within typical bandwidths and ranges \cite{chatterjee2025unifi}, resulting in two distinct designs: GFMI-A and GFMI-B. In Case-A, Bus~36 is equipped with GFMI-A - a compliant GFMI with strong VSBI characteristics. In Case-B, Bus~36 is equipped with GFMI-B - a non-compliant GFMI with weak VSBI characteristics. Fig.~\ref{fig:matlab_compliance} illustrates the frequency-domain compliance assessment of both GFMI designs against ERCOT's reactive power compliance criteria. It can be observed that GFMI-A satisfies ERCOT's $Q(s)/V(s)$ compliance criterion, whereas GFMI-B marginally fails to satisfy the same criterion.

\begin{figure}[htbp]
    \centering
    \includegraphics[width=0.9\linewidth]{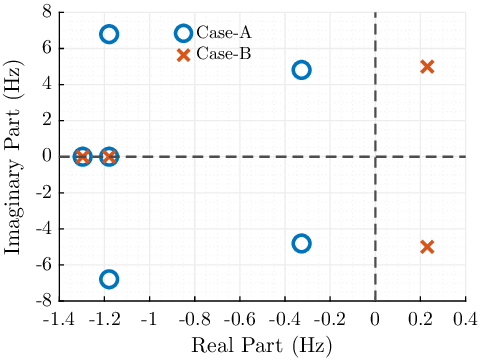}
    \caption{Critical eigenvalues of the modified IEEE 39-bus system for Case-A (compliant GFMI) and Case-B (non-compliant GFMI). The 4.8 Hz mode is stable in Case-A but shifts to the right-half plane in Case-B, demonstrating the impact of VSBI strength on small-signal stability.}
    \label{fig:eigen}
\end{figure}

System-level modal analysis of Case-A (with GFMI-A at Bus~36) and Case-B (with GFMI-B at Bus~36) is presented next. Fig.~\ref{fig:eigen} illustrates the critical eigenvalues for both cases. It can be observed that Case-A, with GFMI-A at Bus~36 (exhibiting stronger VSBI characteristics), has all modes in the left-half plane and is therefore stable. The 4.8~Hz mode is the most critical oscillatory mode ($\lambda_i$), with a damping ratio of 6.77\%. In contrast, Case-B, with GFMI-B at Bus~36 (with weak VSBI characteristics), is unstable due to a right-half-plane 4.8~Hz mode with a negative damping ratio of $-4.61$\%.

\begin{figure}[htbp]
    \centering
    \includegraphics[width=0.9\linewidth]{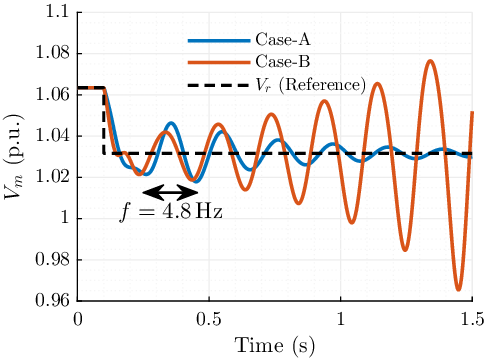}
    \caption{Time-domain verification of the critical 4.8 Hz mode following a 3\% voltage-reference step at Bus 36. The compliant GFMI-A yields a well-damped response, while the non-compliant GFMI-B produces growing oscillations, leading to instability.}
    \label{fig:time_dom}
\end{figure}

The time-domain manifestation of the 4.8~Hz oscillatory mode is illustrated in Fig.~\ref{fig:time_dom}. For a 3\% step reduction in the voltage-magnitude reference of the GFMI at Bus~36, the measured voltage response shows that Case-A (with the compliant GFMI-A) exhibits a well-damped 4.8~Hz oscillatory mode. In contrast, Case-B (with the non-compliant GFMI-B) exhibits a negatively damped 4.8~Hz oscillatory mode, leading to system instability. These time-domain results verify the observations from the system-level modal analysis.

\begin{figure}[htbp]
    \centering
    \includegraphics[width=\linewidth]{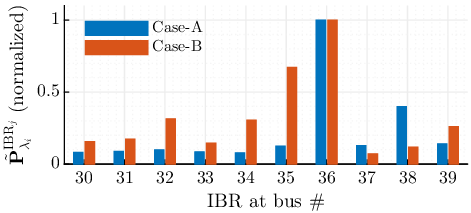}
    \caption{Participation factor $\tilde{\mathbf{P}}_{\lambda_i}^{{IBR}_j}$ of IBR$_j$ in 4.8 oscillatory mode $\lambda_i$. The GFMI at Bus 36 has largest participation, and weakening its VSBI characteristics increases the participation of adjacent GFLIs, leading to instability in Case-B.}
    \label{fig:pf_summed}
\end{figure}

To analyze the participation of each IBR in the critical 4.8~Hz oscillatory mode $\lambda_i$, the system matrix $A^{\mathbf{G}}$ and its associated right and left eigenvectors $(\psi_i, \phi_i)$ are computed. The entries corresponding to the IBR states are extracted, and the contribution of ${IBR}_j$ to mode $\lambda_i$ is quantified using the summed element-wise participation-factor product over the states of that ${IBR}_j$, i.e., $|\phi_i^{{IBR}_j} \odot \psi_i^{{IBR}_j}|$. This yields a scalar participation factor $\mathbf{P}_{\lambda_i}^{{IBR}_j}$. The scalar participation factor $\mathbf{P}_{\lambda_i}^{{IBR}_j}$ is then normalized as $
\tilde{\mathbf{P}}_{\lambda_i}^{{IBR}_j}
=
\mathbf{P}_{\lambda_i}^{{IBR}_j}
\big/
\max_k \mathbf{P}_{\lambda_i}^{{IBR}_k}$. The normalized participation factors of the 4.8~Hz oscillatory mode $\lambda_i$ for each IBR in the system are illustrated in Fig.~\ref{fig:pf_summed}. It is observed that the only generator exhibiting voltage source characteristics in the system, i.e., the GFMI at Bus~36, has the highest participation in the critical 4.8~Hz oscillatory mode, $\lambda_i$. This corroborates the fact that replacing the stronger GFMI-A in Case~A with the weaker GFMI-B in Case~B leads to instability of the mode and, consequently, the system. Moreover, in Case-B, the GFLs at buses 35 and 34, close to the weaker GFMI-B, show much higher participation in the unstable 4.8 Hz mode.

\begin{figure}[htbp]
    \centering
    \includegraphics[width=\linewidth]{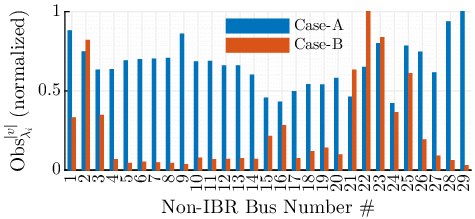}
    \caption{Observability Index ${Obs}^{|v|}_{\lambda_i}$ of the critical 4.8 Hz oscillatory mode $\lambda_i$ in voltage magnitude of non-IBR buses.}
    \label{fig:observability}
\end{figure}

\begin{figure}[htbp]
    \centering
    \includegraphics[width=\linewidth]{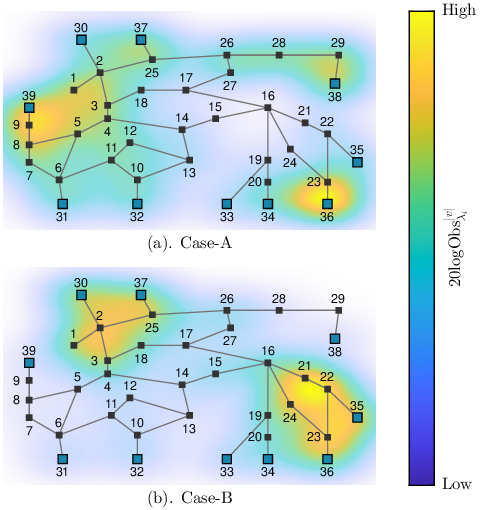}
    \caption{Voltage magnitude observability heatmap indicating vulnerable areas due to critical 4.8 Hz oscillatory mode for: (a). Case-A, and (b). Case-B}
    \label{fig:observability_heat}
\end{figure}

Next, the spatial distribution of the critical 4.8~Hz oscillatory mode $\lambda_i$ is examined. The modal observability ${Obs}_{\lambda_i}$ of mode $\lambda_i$ indicates the grid buses and zones vulnerable to high-amplitude voltage oscillations. Modal observability is calculated as ${Obs}_{\lambda_i} = C^{\mathbf{G}} \psi_i$, where $C^{\mathbf{G}}$ denotes the mapping of system states to bus-voltage outputs. To incorporate the effect of both $d$- and $q$-axis voltage oscillations into a single metric, the observability of the voltage magnitude at each bus is computed as $
{Obs}_{\lambda_i}^{|v|}
=
\sqrt{\left({Obs}_{\lambda_i}^{v_D}\right)^2 + \left({Obs}_{\lambda_i}^{v_Q}\right)^2}$. Fig.~\ref{fig:observability} illustrates the modal observability of the critical 4.8~Hz oscillatory mode at each non-IBR bus in the system, and Fig.~\ref{fig:observability_heat} illustrates an observability heatmap highlighting the most vulnerable areas associated with the critical 4.8~Hz oscillatory mode. It can be observed that, in Case-A, zones near the IBRs at buses 36, 38, and 39 are the most vulnerable, whereas in Case-B, zones near the IBRs at buses 35 and 36 exhibit the highest vulnerability. 


\vspace{-3pt}

\section{Conclusion}
This paper proposed a quantitative frequency-domain compliance framework for assessing the voltage-source-behind-impedance (VSBI) characteristics of grid-forming inverters (GFMIs). The proposed method overcomes the limitations of prevailing time-domain VSBI compliance assessments, which are highly sensitive to measurement-device specifications and dependent on the testing setup. The method is demonstrated by translating the existing time-domain VSBI compliance requirements of AEMO, ERCOT, and MISO into explicit minimum-magnitude Bode-plot compliance envelopes. Modal analysis of the IEEE 39-bus system demonstrates the impact of GFMI VSBI strength and compliance on the small-signal stability of bulk power systems. The application of the method to assess and verify compliance of generic NLR GFMI models in PSCAD using frequency-domain Jacobian scans demonstrates the practicality of the proposed approach. This framework enables system operators to clearly and intuitively visualize GFMI VSBI compliance without reliance on the testing setup or measurement specifications.

\bibliography{IAS_GFM_References}

\end{document}